\documentclass[sigconf]{acmart}

\usepackage{bm}

\newcommand{\vol}{{\rm{vol}}}
\newcommand{\proj}{{\rm{proj}}}

\AtBeginDocument{%
  \providecommand\BibTeX{{%
    \normalfont B\kern-0.5em{\scshape i\kern-0.25em b}\kern-0.8em\TeX}}}

\setcopyright{acmcopyright}
\copyrightyear{2021}
\acmYear{2021}
\acmDOI{}

\acmConference[CAADCPS '21]{the Workshop on Computation-Aware Algorithmic Design for Cyber-Physical Systems: A satellite workshop of the 2021 CPS-IoT Week}{May 18--21, 2021}{Virtual Workshop}
\acmISBN{}

\begin{document}

\title{\sloppy Anytime Ellipsoidal Over-approximation of Forward Reach Sets of Uncertain Linear Systems}

\author{Shadi Haddad}
\affiliation{%
\institution{Department of Applied Mathematics}
  \institution{University of California, Santa Cruz}
  \city{Santa Cruz}
  \state{CA}
  \country{USA}
  \postcode{95064}}
\email{shhaddad@ucsc.edu}

\author{Abhishek Halder}
\affiliation{%
\institution{Department of Applied Mathematics}
  \institution{University of California, Santa Cruz}
  \city{Santa Cruz}
  \state{CA}
  \country{USA}
  \postcode{95064}}
\email{ahalder@ucsc.edu}


\begin{abstract}
Computing tight over-approximation of reach sets of a controlled uncertain dynamical system is a common practice in verification of safety-critical cyber-physical systems (CPS). While several algorithms are available for this purpose, they tend to be computationally demanding in CPS applications since here, the computational resources such as processor availability tend to be scarce, time-varying and difficult to model. A natural idea then is to design ``computation-aware'' algorithms that can dynamically adapt with respect to the processor availability in a provably safe manner. Even though this idea should be applicable in broader context, here we focus on ellipsoidal over-approximations. We demonstrate that the algorithms for ellipsoidal over-approximation of reach sets of uncertain linear systems, are well-suited for anytime implementation in the sense the quality of the over-approximation can be dynamically traded off depending on the computational time available, all the while guaranteeing safety. We give a numerical example to illustrate the idea, and point out possible future directions.
\end{abstract}



\maketitle

\section{Introduction}\label{subsec:Intro}
\begin{sloppypar}
A standard method to verify the performance in safety-critical cyber-physical systems (CPS) is to compute the over-approximation of forward reach sets, i.e., the set of states that the system can reach to at a given time, subject to uncertainties in its initial conditions, control input and unmeasured disturbance. Several numerical toolboxes have been developed for this purpose from different perspectives. For instance, the level set toolbox \cite{mitchell2008flexible} utilizes the fact that the forward reach set is the zero sublevel set of the viscosity solution of certain Hamilton-Jacobi-Bellman partial differential equation associated with the controlled dynamics. There are parametric toolboxes which over-approximate the reach sets using simple geometric shapes such as the ellipsoids \cite{kurzhanskiy2006ellipsoidal} and zonotopes \cite{althoff2015introduction}. There also exist recent works \cite{fan2017d,devonport2020data,devonport2020estimating} for data-driven over-approximation of the reach sets.  
\end{sloppypar}

\begin{sloppypar}
Over-approximating reach sets, even for linear systems, is computationally intensive especially in the presence of time-varying set-valued uncertainties. On the other hand, safety-critical CPS applications have scarce computational resource due to limitations in weight, power, and due to several software concurrently sharing the same hardware. A natural idea then is to design ``computation-aware'' over-approximation algorithms which can dynamically trade-off performance without compromising safety. In particular, one would like to design anytime algorithms \cite{zilberstein1995operational,zilberstein1996using} which provably over-approximate the forward reach sets at any given time but the quality of over-approximation dynamically depends on the computational time available. As more computational time becomes available, the over-approximation becomes ``tighter''.
\end{sloppypar}

\begin{sloppypar}
While anytime algorithms have appeared before in systems-control literature \cite{bhattacharya2004anytime,fontanelli2008anytime,gupta2010control,quevedo2014anytime,pant2015co,liebenwein2018sampling}, their application in parametric over-approximation of the reach set as proposed herein, is new. Specifically, we consider forward reach sets of uncertain linear time-varying systems and point out that the ellipsoidal over-approximation algorithms are particularly suitable for anytime implementation. We summarize the motivations behind ellipsoidal over-approximation in Section \ref{sec:WhyEllipsoids}. The overall computational framework is described in Sec. \ref{sec:Formulation} including the models of dynamics and uncertainties, as well as the anytime ellipsoidal over-approximation algorithm. Sec. \ref{sec:numerics} details a numerical case study to illustrate the ideas. Concluding remarks and future directions are given in Sec. \ref{sec:conclusion}.
\end{sloppypar}

\subsubsection*{Notations} We use $\mathbb{N}$ to denote the set $\{1,2,\hdots\}$, and let $\mathbb{N}_{0}:=\mathbb{N}\cup\{0\}$. The set of $d\times d$ symmetric positive semidefinite (resp. definite) matrices is denoted as $\mathbb{S}_{+}^{d}$ (resp. $\mathbb{S}_{++}^{d}$). The matrix inequalities $\succeq\bm{0}$ (resp. $\succ\bm{0}$) denote positive semidefiniteness (resp. definiteness). A nondegenerate ellipsoid in $d$ dimensions with center vector $\bm{q}\in\mathbb{R}^{d}$ and shape matrix $\bm{Q}\in\mathbb{S}_{++}^{d}$ is given by
\[\mathcal{E}\left(\bm{q},\bm{Q}\right):=\{\bm{y}\in\mathbb{R}^{d}\mid \left(\bm{y}-\bm{q}\right)^{\top}\bm{Q}^{-1}\left(\bm{y}-\bm{q}\right)\leq 1\}.\]
We refer to it as the $\left(\bm{q},\bm{Q}\right)$ ellipsoidal parameterization. As in \cite[Sec. I.2]{halder2018parameterized}, we will also use the $\left(\bm{A}_{0},\bm{b}_{0},c_{0}\right)$ ellipsoidal parameterization encoding the quadratic form:
\[\mathcal{E}\left(\bm{A}_{0},\bm{b}_{0},c_{0}\right):=\{\bm{y}\in\mathbb{R}^{d}\mid \bm{y}^{\top}\bm{A}_{0}\bm{y} + 2\bm{y}^{\top}\bm{b}_{0} + c_{0}\leq 1\}.\]
The $\left(\bm{A}_{0},\bm{b}_{0},c_{0}\right)$ and $\left(\bm{q},\bm{Q}\right)$ parameterizations are related by
\begin{subequations}
\begin{align}
&\bm{A}_{0} = \bm{Q}^{-1}, \quad \bm{b}_{0} = -\bm{Q}^{-1}\bm{q}, \quad c_{0} = \bm{q}^{\top}\bm{Q}^{-1}\bm{q}-1, \label{qQ2Abc}\\
&\bm{Q} = \bm{A}_{0}^{-1}, \quad \bm{q} = -\bm{Q}\bm{b}_{0}.\label{Abc2qQ}	
\end{align}
\label{RelatingParameterizations}	
\end{subequations}
We use $\|\cdot\|_{2}$ to denote the Euclidean 2-norm, $\vol(\cdot)$ to denote the Lebesgue volume, $\lfloor\cdot\rfloor$ to denote the floor operator, and $\mathcal{O}(\cdot)$ for the standard Big-O notation. We use $\bm{0}_{m\times n}$ and $\bm{I}_{n}$ to denote the $m\times n$ zero matrix, and $n\times n$ identity matrx, respectively. We use $\bm{0}_{n}$ for $\bm{0}_{n\times n}$. The symbol $\bm{1}$ denotes a vector of ones of appropriate size. We use shorthands ${\rm{diag}}(\cdot)$ and ${\rm{blkdiag}}(\cdot)$ to denote the diagonal and the block diagonal matrices, respectively. 


\section{Why Ellipsoids}\label{sec:WhyEllipsoids}
The are several reasons why ellipsoids are attractive as a parametric over-approximation primitive.

\begin{enumerate}

\item[(i)] A nondegenerate ellipsoid in $d$ dimensions can be parameterized by $d(d+3)/2$ reals describing its center vector and the shape matrix. Unlike polytopes, this implies fixed parameterization complexity which is useful in CPS context, for example in designing communication protocols where the ellipsoidal descriptions need to be encoded in communication packets. Fixed bit-length parameterization is helpful to reduce the complexity of the communication protocol. 

\item[(ii)] Time-varying ellipsoids naturally model norm bounded uncertainties ubiquitous in systems-control engineering. For example, in vehicular CPS applications, it is natural to represent uncertainties in exogenous disturbance (e.g., wind gust), estimation error and actuation noise as time-varying weighted norm bounds.
	
\end{enumerate}

In the systems-control literature, ellipsoidal over-approximations have been well-investigated in the context of estimation \cite{schweppe1968recursive,witsenhausen1968sets,bertsekas1971recursive,chernous1980guaranteed} and system identification \cite{fogel1979system,norton1987identification,belforte1990parameter,kosut1992set}.


\section{Framework}\label{sec:Formulation}
We next detail the models for dynamics and ellipsoidal set-valued uncertainties, and outline the nature of the computation for ellipsoidal over-approximation of the forward reach set.
\subsection{Models}\label{subsec:Models}
We consider a linear system
\begin{align}
\dot{\bm{x}} = \bm{A}(t)\bm{x} + \bm{B}(t)\bm{u} + \bm{G}(t)\bm{w},
\label{LTVsystem}	
\end{align}
with state $\bm{x}\in\mathbb{R}^{n}$, control input $\bm{u}\in\mathbb{R}^{m}$, and unmeasured disturbance $\bm{w}\in\mathbb{R}^{p}$. The system matrices $\bm{A}(t),\bm{B}(t),\bm{G}(t)$ are assumed to be continuous in time $t$, and are of commensurate dimensions.

Given the set-valued uncertainties in the initial condition $\bm{x}(0)\in\mathcal{X}_{0}$, control $\bm{u}\in\mathcal{U}(t)$, and disturbance $\bm{w}\in\mathcal{W}(t)$, we would like to approximate the forward reach set at time $t$ as
\begin{align}
\mathcal{R}\left(\mathcal{X}_{0},t\right) := \{\bm{x}(t)\in\mathbb{R}^{n}\mid\: &\dot{\bm{x}} = \bm{A}(t)\bm{x} + \bm{B}(t)\bm{u} + \bm{G}(t)\bm{w}, \nonumber\\
&\bm{x}(0)\in\mathcal{X}_{0}, \bm{u}\in\mathcal{U}(t), \bm{w}\in\mathcal{W}(t)\}.
\label{ForwardReachSet}	
\end{align}
We suppose that the set-valued uncertainties are ellipsoidal: $\mathcal{X}_{0}=\mathcal{E}\left(\bm{x}_{0},\bm{X}_{0}\right)$, $\mathcal{U}(t)=\mathcal{E}\left(\bm{u}_{c}(t),\bm{U}(t)\right)$, and $\mathcal{W}(t)=\mathcal{E}\left(\bm{w}_{c}(t),\bm{W}(t)\right)$. In this case, the set $\mathcal{R}\left(\mathcal{X}_{0},t\right)$ is guaranteed to be convex compact.
 
For $k\in\mathbb{N}_{0}$, we consider the prediction horizon $\left[k\Delta t,\allowbreak (k+1)\Delta t\right]$ over which we would like to approximate (\ref{ForwardReachSet}). The \emph{reachable tube} over this prediction horizon is
\begin{align}
\overline{\mathcal{R}}\left(\mathcal{X}_{0},t\right) := \bigcup_{t\in\left[k\Delta t,\allowbreak (k+1)\Delta t\right]}\mathcal{R}\left(\mathcal{X}_{0},t\right).
\label{ReachableTube}	
\end{align}

\subsection{Ellipsoidal Over-approximation of $\mathcal{R}\left(\mathcal{X}_{0},t\right)$}\label{subsec:Ellipsoidal}
We follow the ellipsoidal over-approximation procedure as in \cite{kurzhanskiy2006ellipsoidal}, \cite[Ch. 3]{kurzhanski2014dynamics}. The basic idea is to construct a family of ellipsoids $\{\mathcal{E}\allowbreak\left(\bm{x}_{c}(t),\allowbreak\bm{X}_{i}(t)\right)\}_{i=1}^{N}$ parameterized by unit vectors $\bm{\ell}_{i0}\in\mathbb{R}^{n}$ where $i=1,\hdots,N$. This parameterized family of ellipsoids are constructed such that for any finite $N\in\mathbb{N}$, we have
\begin{align}
\mathcal{R}\left(\mathcal{X}_{0},t\right) \subseteq \widehat{\mathcal{R}}_{N}\left(\mathcal{X}_{0},t\right) := \bigcap_{i=1}^{N} \mathcal{E}\allowbreak\left(\bm{x}_{c}(t),\allowbreak\bm{X}_{i}(t)\right),
\label{Rhat}	
\end{align}
and $\bigcap_{i=1}^{\infty} \mathcal{E}\allowbreak\left(\bm{x}_{c}(t),\allowbreak\bm{X}_{i}(t)\right) = \mathcal{R}\left(\mathcal{X}_{0},t\right)$. Notice that $\widehat{\mathcal{R}}_{N}$ being an intersection of ellipsoids, is guaranteed to be convex, but not an ellipsoid in general.

The center vector $\bm{x}_{c}(t)$ solves the initial value problem (IVP)
\begin{align}
\dot{\bm{x}}_{c} = \bm{A}(t)\bm{x}_{c} + \bm{B}(t)\bm{u}_{c} + \bm{G}(t)\bm{w}_{c}, \quad \bm{x}_{c}(0)=\bm{x}_{0}.
\label{CenterVectorODE}	
\end{align}
Let $\bm{\ell}(t) := \exp\left(-\left(\bm{A}(t)\right)^{\top}t\right)\bm{\ell}_{i0}$, and
\begin{align}
\pi_{i}(t) := \left(\dfrac{\bm{\ell}_{i}^{\top}(t) \bm{B}(t)\bm{U}(t)\bm{B}^{\top}(t) \bm{\ell}_{i}(t)}{\bm{\ell}_{i}^{\top}(t)\bm{X}_{i}(t)\bm{\ell}_{i}(t)}\right)^{1/2}.
\label{defpi}	
\end{align}
Furthermore, define the unit vectors
\begin{align}
\widehat{\bm{v}}_{1i}(t) := \dfrac{\bm{X}_{i}^{1/2}(t)\bm{\ell}_{i}(t)}{\|\bm{X}_{i}^{1/2}(t)\bm{\ell}_{i}(t)\|_{2}}, \quad \widehat{\bm{v}}_{2i}(t) := \dfrac{\bm{G}(t)\bm{W}(t)\bm{G}^{\top}(t)\bm{\ell}_{i}(t)}{\|\bm{G}(t)\bm{W}(t)\bm{G}^{\top}(t)\bm{\ell}_{i}(t)\|_{2}},
\label{defv1v2}	
\end{align}
and let $\bm{S}_{i}(t)$ be an $n\times n$ orthogonal matrix that solves
\begin{align}
\bm{S}_{i}(t)\widehat{\bm{v}}_{2i}(t) = \widehat{\bm{v}}_{1i}(t).
\label{defSi}	
\end{align}
Ref. \cite[Thm. 4.4.4]{kurzhanski2014dynamics} gives an algorithm to compute $\bm{S}_{i}(t)$ in (\ref{defSi}) using $\mathcal{O}(n^{2})$ operations.

With the definitions (\ref{defpi}), (\ref{defv1v2}), (\ref{defSi}) in place, the shape matrices $\bm{X}_{i}(t)$ solve the IVPs
\begin{align}
&\dot{\bm{X}}_{i}(t) = \bm{A}(t)\bm{X}_{i}(t) + \bm{X}_{i}(t)\left(\bm{A}(t)\right)^{\top} \!+ \pi_{i}(t)\bm{X}_{i}(t) + \frac{1}{\pi_{i}(t)}\bm{B}(t)\bm{U}(t)\bm{B}^{\top}\!(t)\nonumber\\
& -\bm{X}_{i}^{1/2}(t)\bm{S}_{i}(t)\bm{G}(t)\bm{W}(t)\bm{G}^{\top}(t) - \bm{G}(t)\bm{W}(t)\bm{G}^{\top}(t)\bm{S}_{i}^{\top}(t)\bm{X}_{i}^{1/2}(t), \nonumber\\
&\bm{X}_{i}(0) = \bm{X}_{0}.
\label{ShapeMatrixODE}	
\end{align}
Solving the IVPs (\ref{CenterVectorODE}) and (\ref{ShapeMatrixODE}) allow us to define $\widehat{\mathcal{R}}_{N}$ in (\ref{Rhat}) that is guaranteed to contain the true reach set $\mathcal{R}$ for any finite $N$. Increasing $N$ results in intersecting more ellipsoids, thus making the outer-approximation tighter. For the derivations of (\ref{CenterVectorODE}) and (\ref{ShapeMatrixODE}), we refer the readers to \cite[Ch. 3]{kurzhanski2014dynamics} and \cite[Part III]{kurzhanskiui1997ellipsoidal}.

Now the question arises how to practically compute/approximate the intersection of a finite number of ellipsoids, which is what $\widehat{\mathcal{R}}_{N}$ is. For parsimony, a natural idea is to compute the minimum volume outer ellipsoid $\mathcal{E}\left(\bm{x}_{c}(t), \bm{X}(t)\right)$, a.k.a. the L\"{o}wner-John ellipsoid \cite{john1948extremum,henk2012lowner},\cite[p. 69]{grotschel1993geometric} containing $\widehat{\mathcal{R}}_{N}$, i.e., to solve
\begin{subequations}
\begin{align}
&\underset{\bm{X}(t)}{\arg\min}\qquad\vol\left(\mathcal{E}\left(\bm{x}_{c}(t),\bm{X}(t)\right)\right)\label{MVOEobj}\\
&\text{subject to} \quad\bigcap_{i=1}^{N} \mathcal{E}\allowbreak\left(\bm{x}_{c}(t),\allowbreak\bm{X}_{i}(t)\right) \subseteq \mathcal{E}\left(\bm{x}_{c}(t),\bm{X}(t)\right).\label{MVOEconstr}	
\end{align}
\label{MVOE}	
\end{subequations}
It is known \cite{john1948extremum}, \cite[Thm. 3.7.1]{ben2001lectures}, \cite[Thm. 3.1.9]{grotschel1993geometric} that the L\"{o}wner-John ellipsoid exists and is unique for any compact convex set, and thus $\bm{X}(t)$ in (\ref{MVOE}) is unique too. However, (\ref{MVOE}) is a semi-infinite programming problem \cite[Ch. 8.4.1]{boyd2004convex} that has no known \emph{exact} semidefinite programming (SDP) reformulation. In fact, verifying (\ref{MVOEconstr}) for $N+1$ given ellipsoids $\{\mathcal{E}\allowbreak\left(\bm{x}_{c}(t),\allowbreak\bm{X}_{i}(t)\right)\}_{i=1}^{N},\mathcal{E}\left(\bm{x}_{c}(t),\bm{X}(t)\right)$, is NP-complete.

Several \emph{suboptimal} reformulations of problem (\ref{MVOE}) are available \cite[Ch. 3.7.2]{boyd1994linear}; one of them is based on the S procedure \cite{yakubovich1971, yakubovich1992nonconvex,polik2007survey} that works well in practice, see e.g.,  \cite[Sec. V]{haddad2021curiousPartI}. We will use this S procedure-based reformulation given by
\begin{subequations}
\begin{align}
&\underset{\widetilde{\bm{A}},\widetilde{\bm{b}},\tau_{1},\hdots,\tau_{N}}{\text{minimize}}\quad\log\det\widetilde{\bm{A}}^{-1}\label{SprocedureObj}\\
&\text{subject to}\qquad\;\; \widetilde{\bm{A}}\succ\bm{0},\\
&\qquad\qquad\qquad\;\;	\tau_{1},\hdots,\tau_{N}\geq 0,\\
&\begin{bmatrix}
\widetilde{\bm{A}} & \widetilde{\bm{b}} & \bm{0}\\
\widetilde{\bm{b}}^{\top} & -1 & \widetilde{\bm{b}}^{\top}\\
\bm{0} & \widetilde{\bm{b}} & -\widetilde{\bm{A}}	
\end{bmatrix} - \displaystyle\sum_{i=1}^{N}\tau_{i}\begin{bmatrix}
\bm{A}_{i} & \bm{b}_{i} & \bm{0}\\
\bm{b}_{i}^{\top} & c_{i} & \bm{0}\\
\bm{0} & \bm{0} & \bm{0}	
\end{bmatrix} \preceq \bm{0}.
\end{align}
\label{Sprocedure}	
\end{subequations}
We note that (\ref{Sprocedure}) is a determinant maximization (max-det) problem subject to linear matrix inequality constraints \cite{vandenberghe1998determinant} for which efficient algorithms are known. Let us denote the optimizer of (\ref{Sprocedure}) as \[\left(\widetilde{\bm{A}}_{\text{opt}},\widetilde{\bm{b}}_{\text{opt}},\tau_{1_{\text{opt}}},\hdots,\tau_{N_{\text{opt}}}\right).\]

Problem (\ref{Sprocedure}) takes the ellipsoids 
\begin{align}
\!\!\bigg\{\mathcal{E}\left(\left(\bm{X}_{i}(t)\right)^{-1}\!,-\left(\bm{X}_{i}(t)\right)^{-1}\!\bm{x}_{c}(t),\bm{x}_{c}^{\top}(t)\left(\bm{X}_{i}(t)\right)^{-1}\!\bm{x}_{c}(t)-1\right)\!\!\bigg\}_{i=1}^{N}\!
\label{InputDataForSprocedure}	
\end{align} 
in $\left(\bm{A}_{0},\bm{b}_{0},c_{0}\right)$ parameterization as input at times $t=k\Delta t$, $k\in\mathbb{N}$. From (\ref{Abc2qQ}), its output in $\left(\bm{q},\bm{Q}\right)$ parameterization is the ellipsoid $\mathcal{E}\left(-\widetilde{\bm{A}}_{\text{opt}}^{-1}\widetilde{\bm{b}}_{\text{opt}},\widetilde{\bm{A}}_{\text{opt}}^{-1}\right)$, which is time-varying since the input data (\ref{InputDataForSprocedure}) is time-varying. If the input ellipsoids are $n$ dimensional, then the max-det problem (\ref{Sprocedure}) has $N + n(n+3)/2$ unknowns.

By construction, the L\"{o}wner-John ellipsoid $\mathcal{E}\left(\bm{x}_{c}(t),\bm{X}(t)\right)$, i.e., the optimal ellipsoid from (\ref{MVOE}), is contained in the optimal ellipsoid obtained from (\ref{Sprocedure}), i.e., in $\mathcal{E}\left(-\widetilde{\bm{A}}_{\text{opt}}^{-1}\widetilde{\bm{b}}_{\text{opt}},\widetilde{\bm{A}}_{\text{opt}}^{-1}\right)$.

\subsection{Parallelization and Projection}

\subsubsection{Parallelizing ellipsoidal propagation}\label{subsubsec:parallelization}
The ellipsoidal overapproximation procedure outlined in Sec. \ref{subsec:Ellipsoidal} involves propagating the center vector $\bm{x}_{c}(t)$ and the shape matrices $\bm{X}_{i}(t)$, followed by solving (\ref{Sprocedure}). Since the solution of the $N+1$ IVPs (\ref{CenterVectorODE}) and (\ref{ShapeMatrixODE}) are independent of each other, they may be run in parallel, if such computing resource is available. Furthermore, the fact that increasing (resp. descreasing) $N$ increases (resp. decreases) the accuracy while guaranteeing the inclusion (\ref{Rhat}), suggests an anytime implementation discussed in Sec. \ref{subsec:Anytime}.

Suppose that the worst-case computational time for propagating $N$ ellipsoids is $t_{\text{propagation}}$. If the worst-case time for solving the IVP (\ref{CenterVectorODE}) is $t_{\text{center}}$, and the same for solving a single instance of the IVP (\ref{ShapeMatrixODE}) is $t_{\text{shape}}$, then $t_{\text{propagation}} = \max\{t_{\text{center}},\allowbreak t_{\text{shape}}\}$ provided parallel computing resource is available. If no parallel computing is available, then $t_{\text{propagation}}=t_{\text{center}}+Nt_{\text{shape}}$. We suppose that $t_{\text{center}}$ and $t_{\text{propagation}}$ are known beforehand (based on the IVP solver used).

\subsubsection{Projection}\label{subsubsec:projection} It is often desired to over-approximate the reach sets of a subset of states. For example, in vehicular CPS applications such as unmanned aerial systems (UAS) and automated driving, ensuring real-time collision avoidance and safe separation amounts to checking distances between reach sets (or over-approximations thereof) in respective position coordinates; e.g., $(x,y,z)$ position coordinates for UAS applications, and $(x,y)$ position coordinates in automated driving applications. 

Notice that since the dynamics remain coupled, the ellipsoidal propagation in Sec. \ref{subsec:Ellipsoidal} need to be done in the original state space $\mathbb{R}^{n}$. However, some computational savings is possible if one is only interested in approximating the reach sets of a subset of states. In such cases, instead of solving the max-det problem (\ref{Sprocedure}) over $n$ dimensional ellipsoids, one may project the propagated ellipsoids on the subset of states of interest, and then solve (\ref{Sprocedure}) over those smaller dimensional ellipsoids, resulting in a lower dimensional convex problem. To justify projection \emph{before} solving (\ref{Sprocedure}), denote $\proj(\cdot)$ as the suitable projection map. Also, let $\mathcal{E}_{\rm{LJ}}(\cdot)$ as the L\"{o}wner-John operator, i.e., a set-valued operator that takes a compact set and returns its unique minimum volume outer ellipsoid. We appeal to the following relations:
\begin{subequations}
\begin{align}
&\!\!\!\!\proj\left(\mathcal{E}_{\rm{LJ}}\left(\bigcap_{i=1}^{N}\mathcal{E}\left(\bm{x}_{c}(t),\bm{X}_{i}(t)\right)\right)\right) = \mathcal{E}_{\rm{LJ}}\left(\proj\left(\bigcap_{i=1}^{N}\mathcal{E}\left(\bm{x}_{c}(t),\bm{X}_{i}(t)\right)\right)\right) \label{LJcommutesWithLinear}\\
&\!\!\!\!\subseteq \mathcal{E}_{\rm{LJ}}\left(\bigcap_{i=1}^{N}\proj\left(\mathcal{E}\left(\bm{x}_{c}(t),\bm{X}_{i}(t)\right)\right)\right) \subseteq \;\stackrel{\text{\normalsize{minimizer of (\ref{Sprocedure}) with}}}{\text{input } \proj(\cdot) \text{ of (\ref{InputDataForSprocedure}).}} \label{Weakening}	
\end{align}	
\label{WhyProjectBeforeOptimize}
\end{subequations}

The equality in (\ref{LJcommutesWithLinear}) holds because the operator $\mathcal{E}_{\rm{LJ}}(\cdot)$ commutes with any linear map \cite[Ch. 8.4.3]{boyd2004convex}. In (\ref{Weakening}), the first set inclusion follows from the general fact that any transformation of intersection is included in the intersection of that transformation. The last set inclusion in (\ref{Weakening}) holds by construction, i.e., because the minimizing ellipsoid of (\ref{Sprocedure}) is a superset of that of the (\ref{MVOE}) for an arbitrary set of input ellipsoids.

We note that projecting the $n$-dimensional ellipsoid to the appropriate axis-aligned subspace amounts to simply extracting the corresponding center subvectors and shape submatrices from the full-dimensional center vectors and shape matrices. The propagation and projection can be parallelized (across unit vectors $\{\bm{\ell}_{i0}\}_{i=1}^{N}$) if such computing resource is available.

\subsection{Anytime Computation}\label{subsec:Anytime}
For $k\in\mathbb{N}_{0}$, suppose that at the instance $t=k\Delta t$,  we have $t_{\text{available}} < \Delta t$ time available to compute an over-approximation of the reach set $\mathcal{R}\left(\mathcal{X}_{0},t=(k+1)\Delta t\right)$. The prediction horizon length $\Delta t$ need not be small. The time $t_{\text{available}}$ will be governed by the processor availability, and may only be known at the instance $t=k\Delta t$. In general, $t_{\text{available}}$ depends on other software running concurrently on the CPS platform, and can have significant variability. Stochastic processor availability models (e.g., i.i.d., Markovian) have appeared before in the anytime control literature \cite{gupta2010control,quevedo2014anytime}.

Recall from Sec. \ref{subsubsec:parallelization} that $t_{\text{propagation}}$ is the worst-case computational time for ellipsoidal propagation. In case any projection on subset of states is performed, we ignore the associated small computational time in extracting the subvectors and submatrices. Suppose $t_{\text{opt}}$ is the worst-case computational time for solving (\ref{Sprocedure}), which has polynomial dependence on $N$ \cite{vandenberghe1998determinant}.

\begin{figure}[t]
\centering
\includegraphics[width=\linewidth]{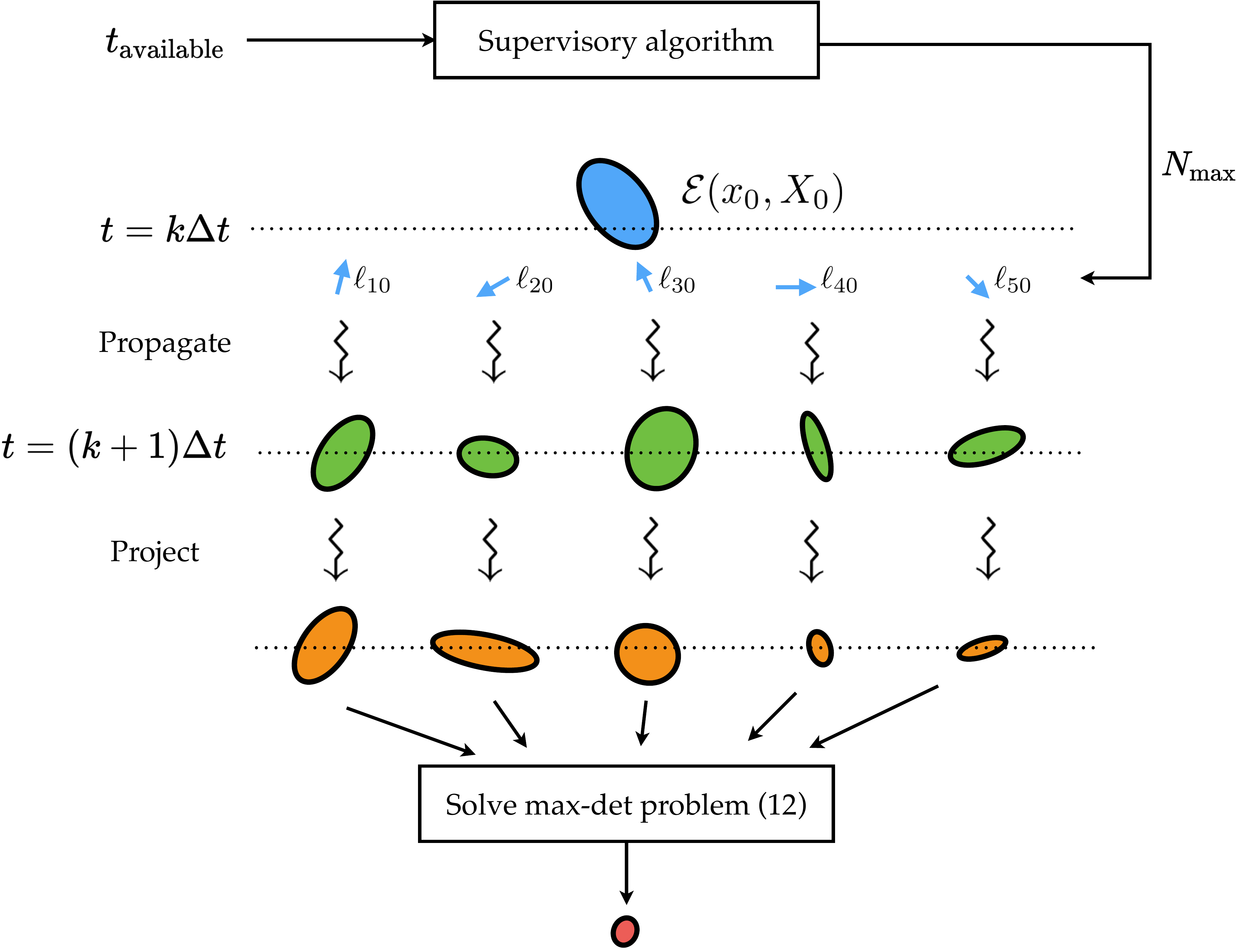}
\caption{A schematic of the proposed anytime computational framework for the ellipsoidal over-approximation of forward reach set at time $t=(k+1)\Delta t$ based on the data at $t=k\Delta t$, $k\in\mathbb{N}_{0}$. Depending on $t_{\text{available}} < \Delta t$, the supervisory algorithm adapts the maximal number of unit vectors $N_{\max}$ (shown here $N_{\max}=5$) to minimize conservatism in over-approximation while preserving safety. The projection step may only be needed when one is interested to compute the reach set over a subset of states. The squiggly arrows denote possible parallelized computation.}
\label{fig:SchematicAnytime}
\end{figure}

Our standing assumption is that $\Delta t$ is large enough to allow the computation in Sec. \ref{subsec:Ellipsoidal} with \emph{at least} $N=1$ (even with no parallel computation), i.e., $t_{\text{center}} + t_{\text{shape}} \leq \Delta t$. Since the total computational time 
\begin{align}
t_{\text{total}} = t_{\text{propagation}} + t_{\text{opt}} = f(N),
\label{TotalCompTime}	
\end{align}
for some nonlinear $f$, a simple way to design the supervisory algorithm shown in Fig. \ref{fig:SchematicAnytime} is to obtain a data-driven estimate $\widehat{f}$ for the function $f$ in (\ref{TotalCompTime}), and then to determine $\widehat{N}$ as the maximal real root of
\begin{align}
t_{\text{available}} = \widehat{f}(\widehat{N}).
\label{Nhat}
\end{align}
As per our assumption, $t_{\text{available}}$ is such that at least $N=1$ is feasible and thus (\ref{Nhat}) has at least one real root. Then $N_{\max} := \lfloor\widehat{N}\rfloor$.
In the numerical results presented in Sec. \ref{sec:numerics}, we computed $\widehat{f}$ using polynomial regression.

The computation for $N_{\max} = 1$ involves single ellipsoidal propagation, and no optimization.

\section{Numerical Simulations}\label{sec:numerics}
\begin{figure}[t]
\centering
\includegraphics[width=\linewidth]{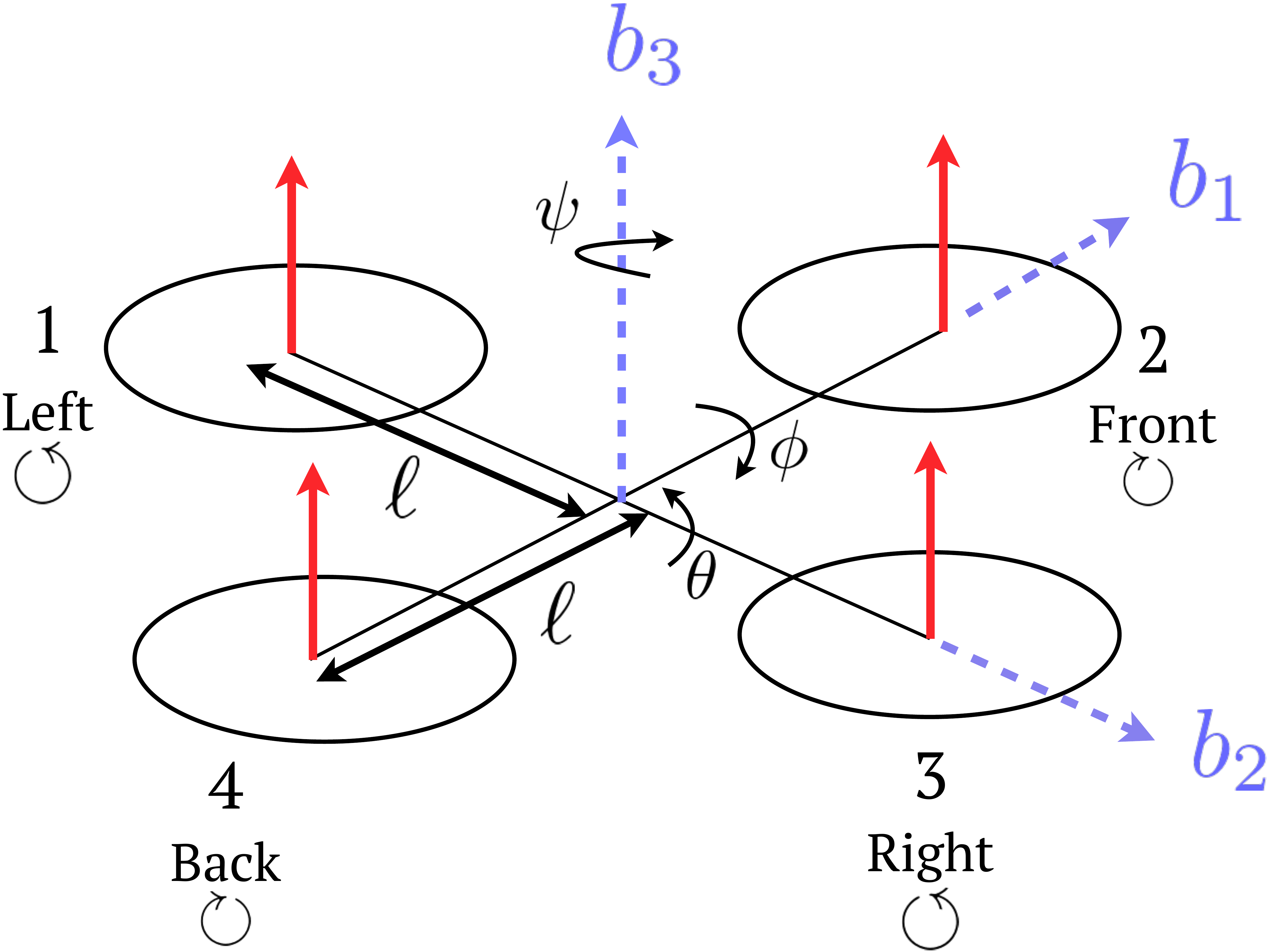}
\caption{A schematic of the rotor numbering convention for the quadrotor dynamics in body frame $b_{1}b_{2}b_{3}$. The parameter $\ell$ denotes the arm length. Also shown are the Euler angles $(\phi,\theta,\psi)$.}
\label{fig:quadrotor}
\end{figure}

To illustrate the ideas presented in Sec. \ref{sec:Formulation}, we consider the linearized model of a standard quadrotor dynamics (see Fig. \ref{fig:quadrotor}) with $n=12$ states, $m=4$ inputs, and $p=3$ unmeasured disturbances. The parameters in the model are shown in Table \ref{tab:param}.

The $12\times 1$ state vector $\bm{x}=\left(x,y,z,\phi,\theta,\psi,p,q,r,u,v,w\right)^{\top}$ comprises of the translational positions $(x,y,z)$ [m], the Euler angles $(\phi,\theta,\psi)$ [rad], the translational velocities $(u,v,w)$ [m/s], and the rotational velocities $(p,q,r)$ [rad/s]. For $i=1,\hdots,4$, the rotor angular velocities (in (rad/s)$^{2}$) are $\omega_{i}^{2} = \overline{\omega}_{i}^{2} + u_{i}$, where the \emph{nominal} rotor angular velocities $\left(\overline{\omega}_{1}^{2},\overline{\omega}_{2}^{2},\overline{\omega}_{3}^{2},\overline{\omega}_{4}^{2}\right)$ solve (from equating thrust to weight and angular torques to zero)
\[\left(\begin{array}{l}
\overline{\omega}_{1}^{2} \\
\overline{\omega}_{2}^{2}\\
\overline{\omega}_{3}^{2}\\
\overline{\omega}_{4}^{2}
\end{array}\right)=\left[\begin{array}{cccc}
c_{\mathrm{T}} & c_{\mathrm{T}} & c_{\mathrm{T}} & c_{\mathrm{T}} \\
\ell c_{\mathrm{T}} & 0 & -\ell c_{\mathrm{T}} & 0 \\
0 & \ell c_{\mathrm{T}} & 0 & -\ell c_{\mathrm{T}} \\
c_{\mathrm{D}} & -c_{\mathrm{D}} & c_{\mathrm{D}} & -c_{\mathrm{D}}
\end{array}\right]^{-1}\left(\begin{array}{c}
m g \\
0 \\
0 \\
0
\end{array}\right).\]
The $4\times 1$ control vector is $\bm{u} = \left(u_1,u_2,u_3,u_4\right)^{\top}$.

\begin{table*}[t]
\centering
\begin{tabular}{|c|l|l|}
\hline
Symbols & Descriptions	& Values [units]\\
\hline\hline
$m$ & mass of quadrotor & 0.468 [kg]\\
\hline
$\ell$ & arm length & 0.225 [m]\\
\hline
${\rm{diag}}\left(J_{xx},J_{yy},J_{zz}\right)
$ & inertia matrix & ${\rm{diag}}\left(5,5,9\right)\times 10^{-3}$ [N$\cdot$m$\cdot$s$^2$]\\
\hline
$c_{\rm{T}}$ & rotor thrust coefficient & $7.2\times 10^{-5}$ [N$\cdot$s$^2$]\\
\hline
$c_{\rm{D}}$ & drag coefficient & $1.1\times 10^{-5}$ [N$\cdot$m$\cdot$s$^2$]\\
\hline
$g$ & acceleration due to gravity & 9.81 [m/s$^2$]\\
\hline
\end{tabular}
\caption{The parameters in the quadrotor model used in Sec. \ref{sec:numerics}.}
\label{tab:param}
\vspace*{-0.2in}
\end{table*}

\begin{sloppypar}
The linearized open-loop model is given by
\begin{align}
\dot{\bm{x}} = \bm{A}\bm{x} + \bm{B}\bm{u} + \bm{G}\bm{w},\quad\bm{G}:= \begin{bmatrix}
 \bm{0}_{6\times 3}\\
 \bm{I}_{3}\\
 \bm{0}_{3}	
 \end{bmatrix},
\label{OpenLoopLTI} 	
\end{align}
i.e., the disturbance $\bm{w}(t)$ models wind gusts acting along the translational acceleration channels, and
\[\bm{A}:=\left[\begin{array}{llll}
\bm{0}_{3} & \bm{0}_{3} & \bm{I}_{3} & \bm{0}_{3} \\
\bm{0}_{3} & \bm{0}_{3} & \bm{0}_{3} & \bm{I}_{3} \\
\bm{0}_{3} & \bm{\Gamma} & \bm{0}_{3} & \bm{0}_{3} \\
\bm{0}_{3} & \bm{0}_{3} & \bm{0}_{3} & \bm{0}_{3}
\end{array}\right], \:\bm{\Gamma}:=\left[\begin{array}{ccc}
0 & -g & 0 \\
g & 0 & 0 \\
0 & 0 & 0
\end{array}\right],\]
\[\boldsymbol{B}:=\left[\begin{array}{c}
\mathbf{0}_{3 \times 4} \\
\mathbf{0}_{3 \times 4} \\
\mathbf{0}_{2 \times 4} \\
\frac{c_{\mathrm{T}}}{m} \mathbf{1}_{1 \times 4} \\
\boldsymbol{\Lambda}_{3 \times 4}
\end{array}\right], \:\boldsymbol{\Lambda}_{3 \times 4}:=\left[\begin{array}{cccc}
\frac{\ell c_{\mathrm{T}}}{J_{x x}} & 0 & -\frac{\ell c_{\mathrm{T}}}{J_{x x}} & 0 \\
& & &\\
0 & \frac{\ell c_{\mathrm{T}}}{J_{y y}} & 0 & -\frac{\ell c_{\mathrm{T}}}{J_{y y}} \\
& & &\\
\frac{c_{\mathrm{D}}}{J_{z z}} & -\frac{c_{\mathrm{D}}}{J_{z z}} & \frac{c_{\mathrm{D}}}{J_{z z}} & -\frac{c_{\mathrm{D}}}{J_{z z}}
\end{array}\right].\]
We close the loop around (\ref{OpenLoopLTI}) using a finite horizon LQR controller 
\begin{align}
\bm{u}(\cdot,t) = \bm{K}(t)\left(\cdot\right) + \bm{u}_{\text{feedforward}}(t)
\label{LQR}	
\end{align}
synthesized to track desired path $(x_{d}(t),y_{d}(t),z_{d}(t))\equiv (\cos t, \sin t, t)$. In the quadratic cost function, we used the state cost weight matrix $\bm{Q}={\rm{blkdiag}}\left(1000\bm{I}_{3},{\rm{diag}}\left(1, 1, 10\right),\bm{I}_{6}\right)$, the control cost weight matrix $\bm{R}=0.1\bm{I}_{4}$, and the terminal cost weight matrix $\bm{M}={\rm{blkdiag}}\left(1000\bm{I}_{3},\bm{I}_{9}\right)$. As is well known (see e.g., \cite[Ch. 4]{anderson2007optimal}), the feedback gain $\bm{K}(t)=-\bm{R}^{-1}\bm{B}^{\top}\bm{P}(t)$ where $\bm{P}(t)$ solves the associated Riccati matrix ODE with terminal condition depending on $\bm{M}$, and that $\bm{u}_{\text{feedforward}}(t) = \bm{R}^{-1}\bm{B}^{\top}\bm{v}(t)$ where $\bm{v}(t)$ solves a vector ODE with terminal condition also depending on the matrix $\bm{M}$.
\end{sloppypar}

\begin{sloppypar}
We suppose that the controller (\ref{LQR}) acts on imperfect state estimate $\widehat{\bm{x}}(t)$ with underestimation error $\bm{\xi}(t):=\bm{x}(t)-\widehat{\bm{x}}(t)$. Letting 
\[\bm{A}_{\rm{cl}}:=\bm{A}+\bm{B}\bm{K}(t), \quad \bm{B}_{\rm{cl}}:= \bm{B}\bm{R}^{-1}\bm{B}^{\top},\quad\bm{\eta}(t):=\bm{P}(t)\bm{\xi}(t)+\bm{v}(t),\]
the closed-loop dynamics can then be written as the linear time-varying system
\begin{align}
\dot{\bm{x}} = \bm{A}_{\rm{cl}}(t)\bm{x} + \bm{B}_{\rm{cl}}\bm{\eta} + \bm{G}\bm{w}. 
\label{LTVclosedloop}	
\end{align}
We suppose that the estimation error $\bm{\xi}(t)\in\mathcal{E}\left(\bm{0}_{12\times 1},\bm{E}(t)\right)$ for known matrices $\bm{E}(t)$ which are $\succ\bm{0}$ at all $t$ and continuous in $t$. Consequently, $\bm{\eta}(t)\in\mathcal{E}\left(\bm{v}(t),\bm{V}(t)\right)$ with $\bm{V}(t):=\bm{P}(t)\bm{E}(t)\bm{P}^{\top}(t)$. Furthermore, $\bm{x}(0)\in\mathcal{E}\left(\bm{x}_{0},\bm{X}_{0}\right)$, $\bm{w}(t)\in\mathcal{E}\left(\bm{w}_{c}(t),\bm{W}(t)\right)$.
\end{sloppypar}

\begin{sloppypar}
We followed the framework in Sec. \ref{sec:Formulation} to propagate the ellipsoidal uncertainties in $12$ dimensional state space and then projected the same in the first three coordinates to obtain the ellipsoidal reach set over-approximation in $(x,y,z)$. We used $\bm{x}_{0}=(1,\bm{0}_{1\times 11})^{\top}$, $\bm{X}_{0}= {\rm{diag}}\left(0.8147,
    0.4854,
    0.7431,
    0.0344,
    0.6551,
    0.9593,
    0.6160,
    \right.$ $\left.
  0.0540, 
   0.1656,
    0.9961,
    0.4314,
    0.5132\right)^{\top}$, $\bm{E}(t)\equiv\bm{I}_{12}$, $\bm{w}_{c}(t)=\left(\cos t, \sin t, \cos t\right)^{\top}$, and $\bm{W}(t)\equiv 0.01\bm{I}_{3}$. All our simulations were done in MATLAB with (\ref{Sprocedure}) solved via \texttt{cvx}.
\end{sloppypar}

\begin{figure}[t]
\centering
\includegraphics[width=0.95\linewidth]{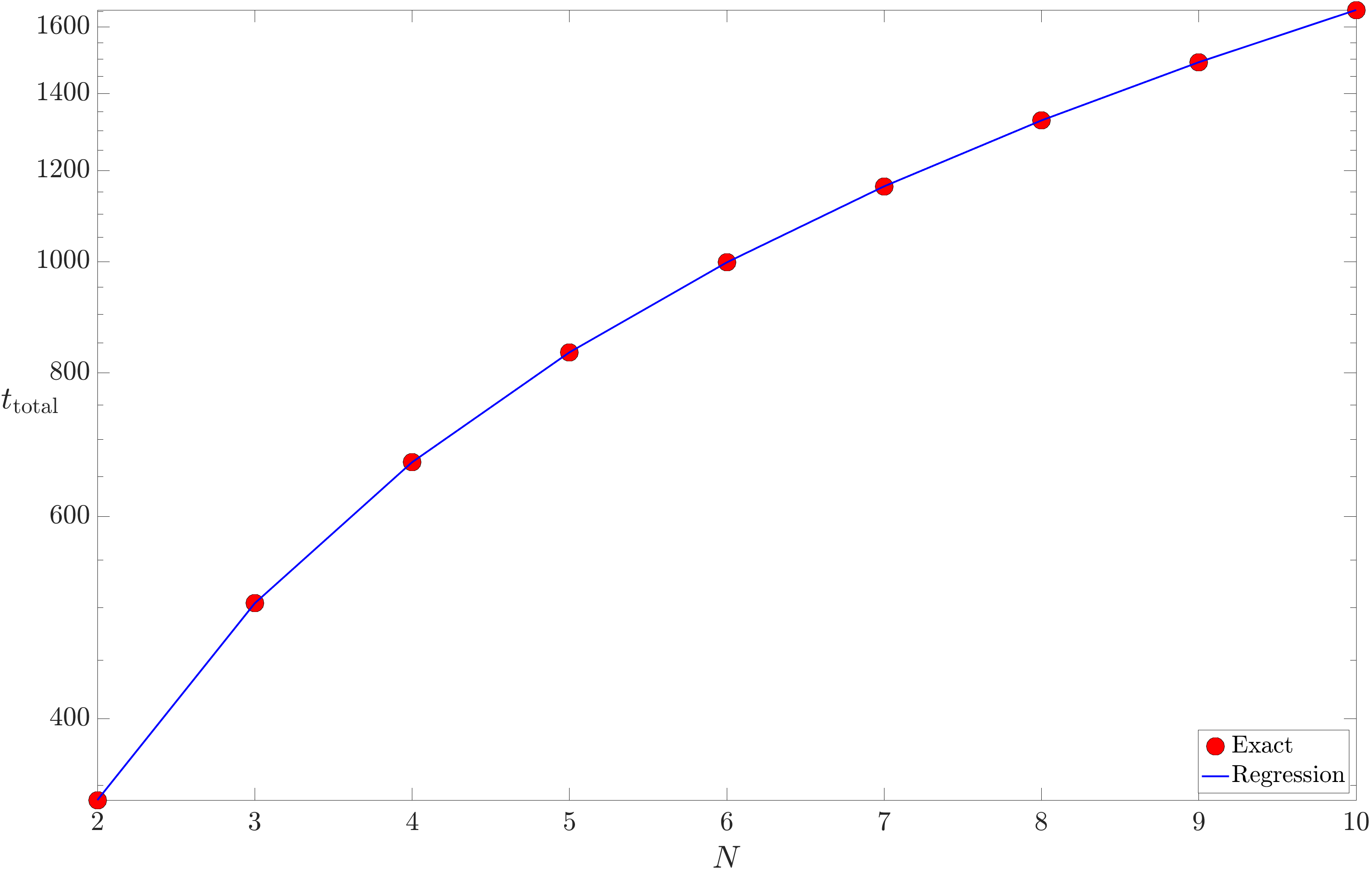}
\caption{Data-driven fourth degree polynomial regression (least square) estimate $\widehat{f}$ for (\ref{TotalCompTime}). The vertical axis is in [s].}
\label{fig:fhat}
\end{figure}

\begin{figure}[t]
\centering
\includegraphics[width=0.95\linewidth]{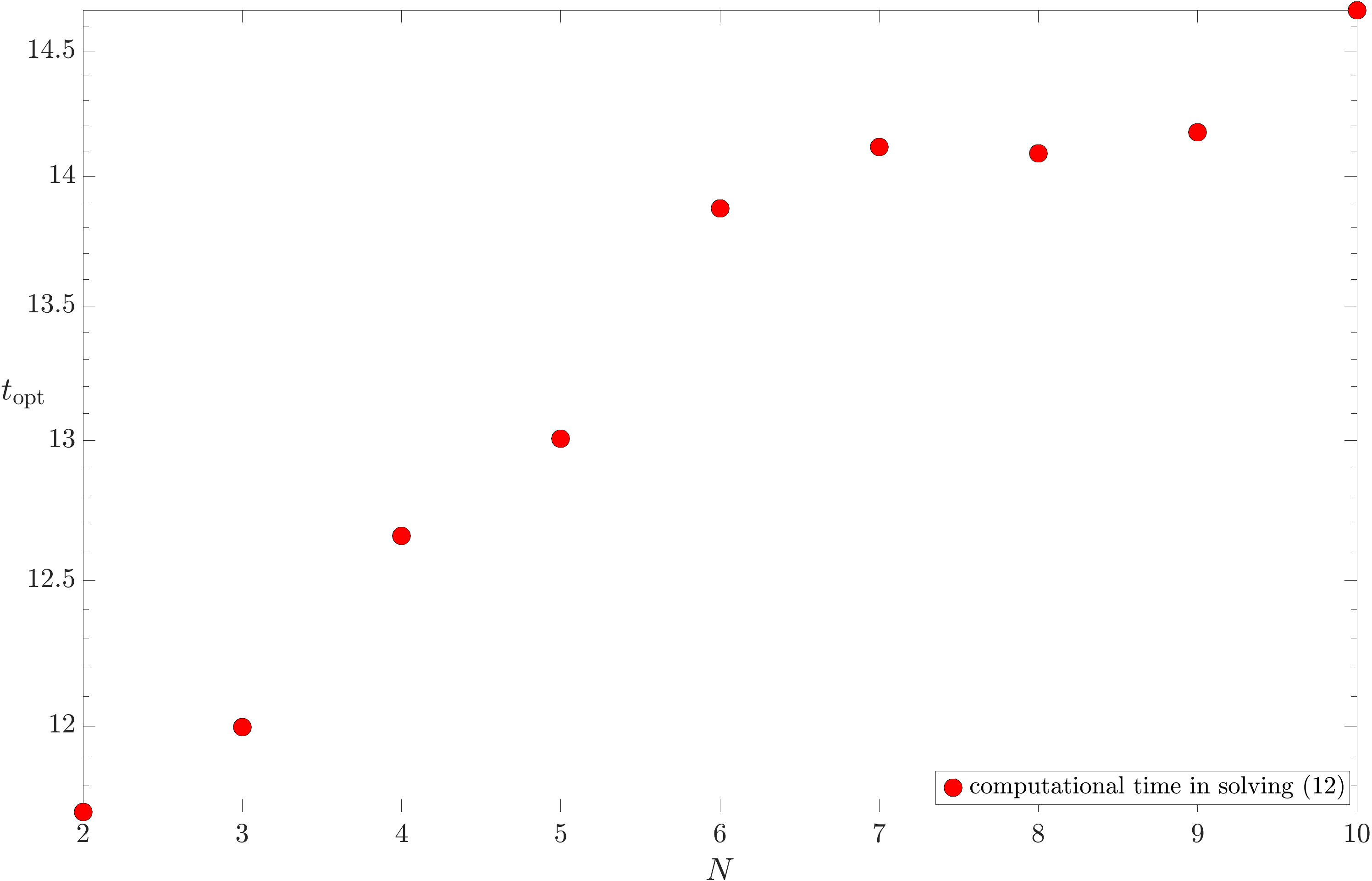}
\caption{Computational times [s] for solving (\ref{Sprocedure}) against $N$.}
\label{fig:opttime}
\end{figure}

To design the supervisory algorithm shown in Fig. \ref{fig:SchematicAnytime} for adapting $N_{\max}$, we used a fourth degree polynomial regression to estimate $f$ in (\ref{TotalCompTime}). The corresponding least square estimate is depicted in Fig. \ref{fig:fhat}. Fig. \ref{fig:opttime} reveals that $t_{\text{opt}} << t_{\text{propagation}}$ for our simulation, i.e., $t_{\text{total}}$ is dominated by the time needed to solve the IVPs.

\begin{figure}[t]
\centering
\includegraphics[width=\linewidth]{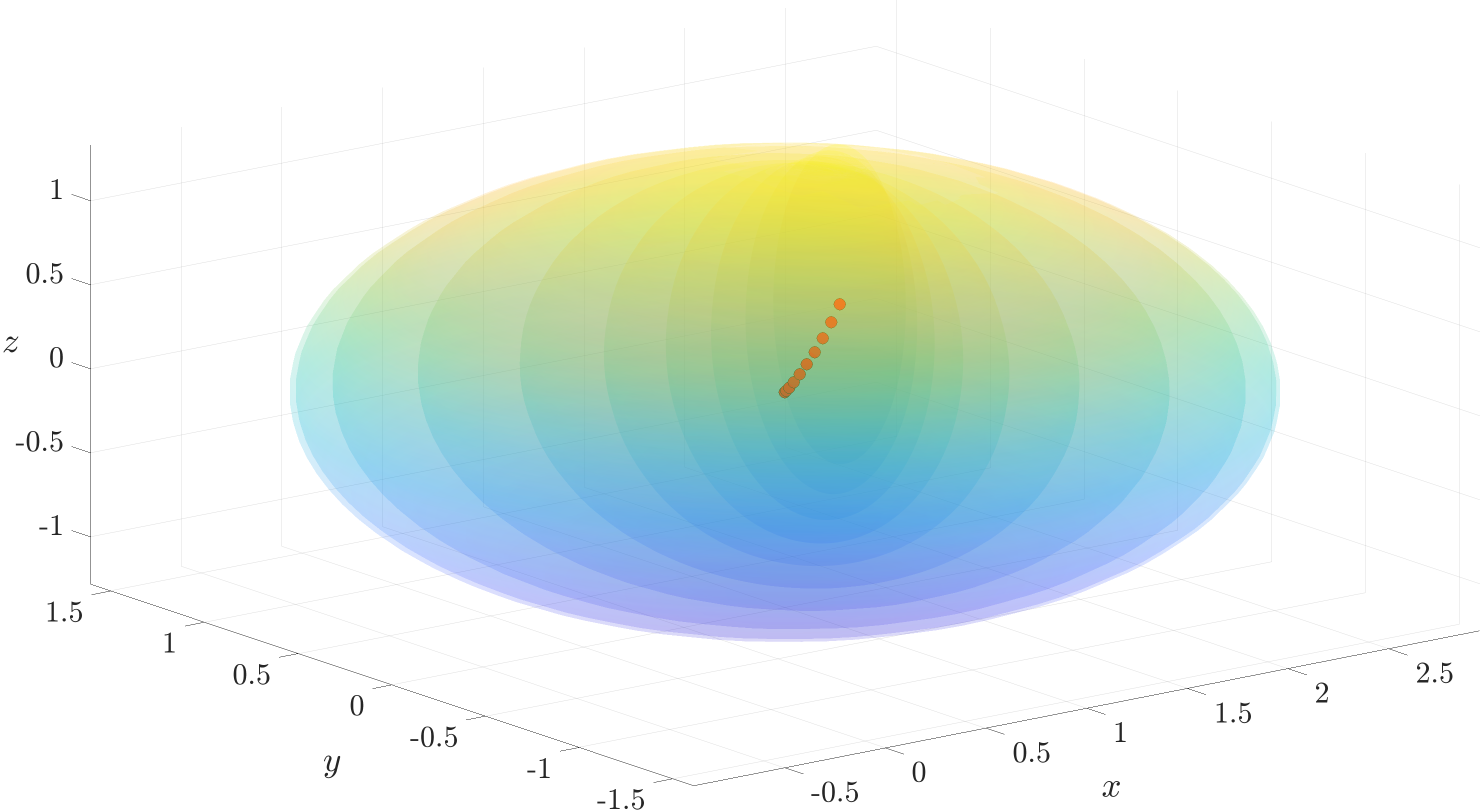}
\caption{The ellipsoidal outer-approximations of the reach set in $(x,y,z)$ for the simulation case study in Sec. \ref{sec:numerics} with $N_{\max}=10$. Shown here are 10 equi-spaced time snapshots superimposed for $t\in[0,1]$. The red dots show the centers of the respective ellipsoidal over-approximations.}
\label{fig:EllipsoidalOverApproxN10}
\end{figure}

In Fig. \ref{fig:EllipsoidalOverApproxN10}, we show the ellipsoidal over-approximations for reach sets in the position coordinates $(x,y,z)$ for $t\in[0,1]$, obtained using the proposed framework.


\section{Conclusions and Future Work}\label{sec:conclusion}
\begin{sloppypar}
We outlined an anytime ellipsoidal over-approximation framework for the forward reach sets of an uncertain linear system with ellipsoidal set-valued uncertainties. Our main intent was to point out that the existing ellipsoidal over-approximation results are well-positioned for anytime implementation, thereby opening up the possibility to deploy them for safety-critical CPS applications in a manner that not only acknowledges the limited computational resource in these settings, but dynamically adapts its performance depending on processor availability without sacrificing safety. We provided a numerical case study to elucidate the ideas.

Several avenues of future work remain open. For example, instead of regression, one may design the supervisory algorithm for computing $N_{\max}$ via online learning. It may also be interesting to analyze the performance of these anytime algorithms under stochastic processor availability models, as was done in control settings \cite{gupta2010control,quevedo2014anytime,fontanelli2008anytime}. One may also be interested to design anytime algorithms for other parametric over-approximation algorithms, e.g., using zonotopes \cite{althoff2015introduction}. 
\end{sloppypar}

\begin{acks}
This research was partially supported by a 2018 Faculty Research Grant by the Committee of Research from the University of California, Santa Cruz, a 2018 Seed Fund Award from CITRIS and the Banatao Institute at the University of California, a 2019 Ford University Research Project, and a Chancellor's Fellowship from the University of California, Santa Cruz.
\end{acks}


\bibliographystyle{ACM-Reference-Format}
\bibliography{CAADCPS2021-ShadiAbhishek.bib}


%
%
%
%
%
%
%

\end{document}